\documentclass[epj]{svjour}
\newcommand{\be}{\begin{equation}}    
\newcommand{\ee}{\end{equation}}
\newcommand{\ba}{\begin{eqnarray}}
\newcommand{\ea}{\end{eqnarray}}

\usepackage{graphicx}  
\begin{document}
\title{Relevance of the 1:1 resonance in galactic dynamics}
\author{A.~Marchesiello\inst{1}
\and
G.~Pucacco\inst{2}\thanks{pucacco@roma2.infn.it.}}
\institute{Dipartimento di Scienze di Base e Applicate per l'Ingegneria -- Universit\`a di Roma ``la Sapienza",
Via Antonio Scarpa, 16 - 00161 Roma  
\and Dipartimento di Fisica -- Universit\`a di Roma ``Tor Vergata" and INFN -- Sezione di Roma Tor Vergata, Via della Ricerca Scientifica, 1 - 00133 Roma}

\date{Received: date / Revised version: date}
%

\abstract{This paper aims to illustrate the applications of resonant Hamiltonian normal forms to some problems of galactic dynamics. We detail the construction of the 1:1 resonant normal form corresponding to a wide class of potentials with self-similar elliptical equi-potentials and apply it to investigate relevant features of the orbit structure of the system. We show how to compute the bifurcation of the main periodic orbits in the symmetry planes of a triaxial ellipsoid and in the meridional plane of an axi-symmetric spheroid and briefly discuss how to refine these results with higher-order approaches.
}
\maketitle

\section{Description}
Nonlinear dynamics has often profited, since the pioneering works of Jeans, Lindblad, etc., of the progress in galactic dynamics. The interplay between the two disciplines has gradually established a common body of methods and results, as much as in the case of the interplay with celestial mechanics and is well documented in the literature~\cite{bp,ber,conto}.  Nowadays, advanced techniques of mathematical physics provide rigorous analysis of the main phenomena occurring in non-integrable dynamical systems potentially useful in galactic dynamics. However, the language of these modern methods often prevent a direct use of those results to applied fields like astrophysics.

Aim of this work is to show how to apply modern perturbative methods in analytical mechanics to provide a simple and comprehensive description of a common phenomenon in the dynamics of stellar systems: the onset of instability of a given motion due to a nonlinear resonance and the bifurcation of a new solution. We limit our analysis to the case in which the two involved frequencies, that of the perturbed solution and that of the `perturbation', happen to become equal as functions in a suitable parameter space. This is called a `1:1 resonance' and plays a distinguished role in several concrete cases. The approach we use is that of mimicking the dynamics of the original physical system, that cannot be exactly solved, with an approximating system which is exactly solvable. The original system is assumed to be a generic non-integrable Hamiltonian system: the mimicking system is an integrable resonant Hamiltonian `normal form'. We give explicit formulae to construct the normal form of a fairly generic class of systems and show how to use it to gather a general view of the phase space structure around the resonance. We prove that, in addition to a complete qualitative understanding, the theory also provides quite accurate {\it quantitative} predictions, even at the simplest level of the procedure. We finally discuss how to refine the predictions by going to higher-orders and how to generalize the method in a wider context.

The plane of the paper is as follows: in section 2 we introduce the procedure to construct the approximating integrable system by recalling the method of the {\it Lie transform} \cite{GS,gior}; in sections 3 and 4 we apply this approach to investigate some aspects of the dynamics in a symmetry plane of a triaxial ellipsoid \cite{bbp} and in the meridional plane of an axisymmetric spheroid \cite{p09}, obtaining first-order estimates of the bifurcation thresholds of the 1:1 periodic orbits; in section 5 we discuss further developments and hints for other applications.

\section{Normal forms}

The general setting in which we cast the dynamics of the physical systems we are interested in is that of a natural Hamiltonian in two degrees of freedom in which the potential is assumed to be expanded as a truncated series (in the coordinates $\vec{q} = (x,y) $) of the type
\be\label{potes}
V^{(N)} (\vec{q};\vec{\alpha}) = 
\sum_{k=0}^{N} \sum_{j=0}^{k} C_{(j,k-j)}(\vec{\alpha})x^{j}y^{k-j},
\ee
where the truncation order $N$ and the parameters $\vec{\alpha}$ are determined by the problem under study. We observe that for our `galactic' applications $N=4$ is already enough to describe the most relevant examples and moreover we will usually assume reflection symmetry in one or both the coordinate axes. The Hamiltonian 
\be\label{HaS}
H = \frac12 |\vec{p}|^{2} + V^{(N)} (\vec{q};\vec{\alpha}) ,\ee
is in the form of a power series and is therefore naturally apt to be treated in a perturbative way as a non-linear oscillator system. We construct a `normal form'~\cite{bp}, namely a new Hamiltonian series which is an integrable approximation of the original one, suitably devised to catch its most relevant orbital features. 

The normal form is `non-resonant' when the two harmonic frequencies 
\be \label{uf}
\omega_1 \doteq  \sqrt{2 C_{(2,0)}} ,  \quad
\omega_2 \doteq  \sqrt{2 C_{(0,2)}}, \ee
are generically non-commensurable: in this case we get explicit formulas for actions and frequencies of the box orbits parented by the $x$- and $y$-axis periodic orbits. A `resonant' normal form is instead assembled by assuming from the start an integer value for the ratio of the harmonic frequencies and including in the new Hamiltonian terms depending on the corresponding resonant combination of the angles. This possibility might be thought to be exceptional: it is instead {\it almost the rule} because, even if the unperturbed system is non-resonant with a certain real value 
\be \label{ratio} \rho = \omega_1/\omega_2 \ee
of the frequency ratio, the non-linear coupling between the degrees of freedom induced by the perturbation determines a `passage through resonance' with a commensurability ratio, say $m_1/m_2$ with $ m_1,m_2 \in \mathbf N$, corresponding to the local ratio of oscillations in the two degrees of freedom. This in turn is responsible of the birth of new orbit families bifurcating from the normal modes or from lower-order resonances. 

We assume that our system is such that the ratio (\ref{ratio}) is not far from a rational value and then approximate it by introducing a small `detuning' $\delta$ so that
\be\label{DET}
\rho = m_1/m_2 + \delta\ee
and proceed like if the unperturbed harmonic part would be in exact $m_1$:$m_2$ resonance by putting the remaining part inside the `perturbation'. We speak of a {\it detuned} ($m_1$:$m_2$) {\it resonance}, implying with this a trick to deceive the formally correct but ineffective approach based on non-resonant generic frequency ratios.

Resonant normal forms for the Hamiltonian system corresponding to (\ref{HaS}) are constructed with standard methods: here we are going to apply the method based on the Lie transform \cite{bp}. The usual approach in Hamiltonian perturbation theory is to find a generating function in such a way to construct a `simpler' Hamiltonian: the first successful algorithm devoted to a resonant system was precisely introduced to treat the problem of the `third integral' in galactic dynamics \cite{gu}. The method based on the Lie transform offers several conceptual and technical advantages \cite{bp,gior} with respect to the original approach. The idea behind the method is that of seeing the canonical transformation as a `flow' along the Hamiltonian vector field associated to the generating function. In spite of this apparently exotic statement, it turns out that this approach is that best suited in implementing recursive algorithm working with series of functions. In the following we provide the clue on how to set up the method remarking that for our applications only the first steps of the procedure will be necessary.

Let us consider a phase-space function expanded as a power series in the canonical variables
\be\label{gene}
G=G_{0}+G_{1}+G_{2}+...\ee 
We assume that all functions are expanded in polynomial series and we adopt the convention that the `zero order' terms of the series are quadratic homogeneous polynomials and terms of {\it order} {\it n} are polynomials of degree $n+2$. 

To the series (\ref{gene}) is naturally associated the linear differential operator
\be\label{diff}
{\rm e}^{{\cal L}_G} = \sum_k \frac1{k!} {\cal L}_G^k,\ee
whose action on a generic function $F$ is given by the Poisson bracket:
\be
{\cal L}_G F \doteq \{F,G\}.\ee
After a scaling transformation
\be p_{x} \longrightarrow  \sqrt{\omega_1} \, p_{x}, \quad  p_{y} \longrightarrow \sqrt{\omega_2} p_{y}, \quad  
x \longrightarrow x/\sqrt{\omega_1},  \quad
y \longrightarrow y/\sqrt{\omega_2},\ee
the original Hamiltonian system (\ref{HaS}) undergoes a canonical transformation to new variables $
(\vec{P},\vec{Q}),$ such that the new Hamiltonian is
\begin{equation}\label{HK}
     K(\vec{P},\vec{Q})={\rm e}^{{\cal L}_G} H(\vec{p},\vec{q}),\ee
    where both functions are assumed to be in the form of power series 
    \be H=\sum_{n=0}^{N}H_n, \quad K=\sum_{n=0}^{N}K_n.\ee

To construct $K$ starting from $H$ is a recursive procedure exploiting an algorithm based on the Lie transform \cite{bp,gior}. To understand how it works, let us consider the first step of the procedure, namely let us perform a transformation given by a function $G_1$ considered as the first term in the generating function (we can start with the cubic term because $G_{0}$ gives only trivial linear transformations). The general relation (\ref{HK}) takes the form
\begin{equation}\label{HK1}
K_0 + K_1 + ... = (1 + {\cal L}_{G_1} + ...) (H_0 + H_1 + ...).\ee
By equating polynomials of the same degree, we get the system:
\ba
K_0 &=& H_0,\\
K_1 &=& H_1 - {\cal L}_{G_1} H_0,\\
K_2 &=& H_2 - {\cal L}_{G_1} H_1 - {\scriptstyle \frac12}  {\cal L}^2_{G_1} H_0\ea
and equations involving terms of higher degrees. The first equality simply states that the zero-order new Hamiltonian,  
\be\label{Hzero}
K_{0} = \frac12 \left(\omega_1 (P_X^2 + X^2) + \omega_2 (P_Y^2  +Y^2) \right),
\ee
coincides with the zero-order old (unperturbed) one. The second equation has to be solved to find the first order term $K_1$. At this point we are faced with a difficulty: we have {\it one} differential equation involving {\it two} unknown functions, $K_1$ and  $G_1$. To proceed we have to make some decision about the structure the new Hamiltonian must have, that is we have to chose a {\it normal} form for it. We therefore select the new Hamiltonian in such a way that it admit a new integral of motion, that is we take a certain function, say $I$, and impose that
\be\label{NFD}
\{K,I\}=0.
\ee
The usual choice (but not necessarily the only possible) is that of taking
\be\label{UC}
I=H_{0}=K_0
\ee
so that the function (\ref{Hzero}) plays the double role of determining the specific form of the transformation and assuming the status of the second integral of motion. With this choice, the fundamental equation of the chain, that we can also write in the form
\be\label{homo}
K_1 = H_1 -  {\cal L}_{G_1} H_0 = H_1 +  {\cal L}_{H_0} G_1,\ee
is solved with a trick that we illustrate in the following. The action of the operator $ {\cal L}_{H_0}$ on any polynomial function which commutes with $H_0$ (and therefore that has  $H_0$ as an integral of motion) is to kill it, whereas its action on any other polynomial gives a uniquely defined non-vanishing polynomial. We can therefore split the polynomial  $H_1$ appearing in (\ref{homo}) according to the rule
\be
H_1 = H_1^{\rm K} + H_1^{\rm R}\ee
where $H_1^{\rm K}$ is the part which stays in the {\it kernel} of  $ {\cal L}_{H_0}$, that is is such that 
\be
 {\cal L}_{H_0} H_1^{\rm K} \equiv 0,\ee
and 
 $H_1^{\rm R}$ is the part which stays in the {\it range} of  $ {\cal L}_{H_0}$, that is is such that 
\be
 {\cal L}_{H_0} H_1^{\rm R} = R_1,\ee
where $R_1$ is a non-vanishing cubic polynomial. Since our new Hamiltonian, with the choice made, is in normal form if and only if it stays in the kernel of  $ {\cal L}_{H_0}$, we can then {\it solve} eq.(\ref{homo}) by applying the simple prescription:
\be\label{homosol}
K_1 = H_1^{\rm K} , \quad  G_1 = {\cal L}^{-1}_{H_0} H_1^{\rm R}.\ee
We have then constructed the {\it normal form} $K_0 + K_1$ at order 1 and computed the first term of the generating function $G_1$: we can therefore use it in the subsequent equations of the system to compute the terms $K_n, n > 1$ (which are still {\it not} in normal form) and go one step further by expanding eq.(\ref{HK1}) at order two and repeating the procedure to compute $G_2$ and the normal form at order 2 and so forth.

Formally, a more direct way of applying this method is by using `action-angle'--{\it like} variables $\vec{J}, \vec{\theta}$ defined through the transformation
\ba
X &=& \sqrt{2 J_1} \cos \theta_1,\quad
Y  = \sqrt{2 J_2} \cos \theta_2,\label{AAV1}\\
P_X &=& \sqrt{2 J_1} \sin \theta_1,\quad
P_Y = \sqrt{2 J_2} \sin \theta_2.\label{AAV2}\ea 
In this way we have
\be
H_0 = \omega_1 J_1 + \omega_2 J_2, \ee
so that
\be\label{LHAA}
 {\cal L}_{H_0} = \omega_1 \frac{\partial}{\partial \theta_1}  + \omega_2 \frac{\partial}{\partial \theta_2} . \ee
Moreover, a generic polynomial series turns out to be a Fourier series in the angles with coefficients depending on the actions. It is clear that in the non-resonant case, the kernel  of  $ {\cal L}_{H_0}$ is composed only of functions of the actions and this must also be the case for the normal form: in this case we speak of a Birkhoff (or non-resonant) normal form. But if the frequencies are (almost) in the ratio $m_1/m_2$,  
the typical structure of the resonant normal form (truncated when the first resonant term appears) is \cite{conto}
\begin{eqnarray}\label{GNF}
K&=&m_{1} J_1+m_{2} (J_2 +  \delta J_1) + \sum_{k=2}^{m_1+m_2} {\cal P}^{(k)}(J_1,J_2)+\nonumber \\
&&a_{m_1 m_2} J_1^{m_{2}} J_2^{m_{1}} \cos [2(m_{2} \theta_{1}- m_{1} \theta_{2})], 
\end{eqnarray}
where ${\cal P}^{(k)}$ are homogeneous polynomials of degree $k$ whose coefficients may depend on $\delta$ and the constant $a_{m_1 m_2}(\vec{\alpha})$ is the only marker of the resonance. It is easy to check that this is the most general form of a phase-space function of degree $m_1+m_2$ in the actions which stays in the kernel of  $ {\cal L}_{H_0}$ as given by (\ref{LHAA}) if the frequencies are such that, in agreement with assumption (\ref{DET}),
\be
m_1/m_2 =  \omega_1 / \omega_2 +  \delta.\ee

In these variables, the second integral is 
\be\label{cale}
{\cal E}=m_{1} J_1+m_{2} J_2\ee
and the angles appear only in the resonant combination
\be\label{psi}
\psi=m_{2} \theta_{1}- m_{1} \theta_{2}.\ee
For a given resonance, these two statements remain true for arbitrary $N>m_{1}+m_{2}$. Introducing the variable conjugate to $\psi$,
\be\label{calr}
{\cal R}=m_{2} J_1-m_{1} J_2,\ee
the new Hamiltonian can be expressed in the {\it reduced} form $K({\cal R}, \psi; {\cal E},\vec{\alpha})$, that is a family of 1-dof systems in the phase-plane ${\cal R}, \psi$, parametrized by ${\cal E}$ (and $\vec{\alpha}$). 

We recall that, as in (\ref{DET}), the resonant ratio is not exact: as a result we have that the quadratic part can be written in the form
\be
H_0 = \omega_2 \left( \frac{m_1}{m_2} J_1 + J_2 \right) + \omega_2 \delta J_1. \ee
We can then rescale the Hamiltonian with the constant factor $m_2/\omega_2$ and normalize with respect to the exactly resonant quadratic part (\ref{cale}): the small `detuning' term $m_2  \delta J_1$, even if still quadratic in the coordinates, is treated as a higher-order term and inserted into the perturbation $H_1+H_2+...$ Since in practice $\delta$ is related to some morphological parameters of the potential, this trick allows us to considerable extend the range of the analysis. In the applications we aim at illustrating the fundamental case in which the ratio between the integers is simply one. Therefore, we will explicitly compute the 1:1 resonant normal form that, when truncated to the first non zero resonant term, is given by 
\be\label{Kq11}
K  = J_1 + J_2 +  \delta J_1 + a J_1^{2} + b J_2^{2} + J_1J_2 \left(2 c + d \cos [2(\theta_{1}- \theta_{2})] \right),\ee
where the coefficients $a,b,c$ and $d=a_{11}$ are the result of the normalization process and are completely defined in terms of the parameters of the original system.

In the applications below we are interested in the global structure of phase-space, but also the explicit solution of the equations of motion is of great relevance. For a non-resonant normal form, the problem is easily solved: the coefficient $a_{m_1 m_2}$ vanishes and $K$ no longer has terms containing angles. Therefore, the $\vec{J}$ are `true' conserved actions and the solutions are
\be\label{SF}
X (\tau) = \sqrt{2 J_1} \cos \kappa_1 \tau ,\quad
Y (\tau) = \sqrt{2 J_2} \cos (\kappa_2 \tau + \theta_0),\ee
where
\be
\vec{\kappa} = \nabla_{\vec{J}} K\ee
is the frequency vector and $\theta_0$ is a suitable phase shift. In the resonant case, in general it is not possible to write the solutions in closed form. It is true that the dynamics described by the 1-dof Hamiltonian $K({\cal R}, \psi)$ are always integrable, but the solutions cannot be written in terms of elementary functions. However, solutions can still be written down in the case of the {\it main periodic orbits}, for which $\vec{J}, \vec{\theta}$ are true action-angle variables. There are two types of periodic orbits that can be easily identified:
\begin{enumerate}
      \item The {\it normal modes}, for which one of the $\vec{J}$ vanishes.
         
      \item  The periodic orbits {\it in general position}, namely those solutions characterized by a {\it fixed} relation between the two angles, $m_{2} \theta_{1}- m_{1} \theta_{2} \equiv \theta_0$.
       \end{enumerate}
In both cases, it is straightforward to check that the solutions retain a form analogous to Eq.(\ref{SF}) with known expressions of the actions and frequencies in terms of ${\cal E}$ and the parameters $\vec\alpha$ such that $\kappa_1 / \kappa_2 = m_1 / m_2$. By using the generating function Eq.(\ref{gene}), the solutions in terms of standard `physical' coordinates can be recovered (taking into account possible scaling factors) by inverting the canonical transformation. The transformation back to the physical coordinates gives a series of the form
\be
\vec{q} (\tau) = \vec{q}_{1} + \vec{q}_{2} + \vec{q}_{3} +...\ee   
and can be computed by exploiting the properties of the Lie series as above:
\ba
\vec{q}_{1} &=& \vec{Q}, \label{x1}\\
\vec{q}_{2} &=& \{G_{1},\vec{Q}\}, \label{x2} \\
\vec{q}_{3} &=& 
\{G_{2},\vec{Q}\} + {\scriptstyle\frac12} \{G_{1},\{G_{1},\vec{Q}\}\} \label{x3} \ea
and so forth. From a knowledge of the normalized solutions Eq.(\ref{SF}), we can therefore construct \cite{p08} power series approximate solutions of the equations of motion of the original system
\be
\frac{d^{2} \vec{x}}{d \tau^{2}} = -\nabla_{\vec{x}} V^{(N)} (x,y;\vec{\alpha}).\ee

As a rule, normal modes exist on each `energy' surface $K=\widetilde E = (m_2/\omega_2) E$. Periodic orbits in general position exist instead only beyond a certain threshold and we speak of a bifurcation ensuing from a detuned resonance. The bifurcation is usually described by a series expansion of the form
\be\label{detexp}
 \widetilde E = \sum_{k} a_{k} |\delta|^{k},\ee
where the $a_{k}$ are coefficients depending on the order $N$ and the parameters $\vec{\alpha}$. The order of truncation of the series is related to that of the normal form. Eq.(\ref{detexp}) implies that at exact resonance (vanishing detuning) the bifurcation is intrinsic in the system and that, going away from the exact ratio of unperturbed frequencies, gradually increases the threshold value for the bifurcation.  We will see that already a linear relation given by the first order truncation provides a reliable estimate of the threshold values if due care is made of the consistency of series expansions.  

\section{The stability of axial orbits in systems with elliptical equipotentials}

We show how to exploit the above theory in some cases of interest in galactic dynamics. We start with the problem of the stability of axial orbits in triaxial potentials and of the possible existence of additional stable families of periodic orbits. The relevance of such issues in galactic dynamics is for problems like the construction of self-consistent equilibria, the computation of isophotal shapes and velocity ellipsoids, etc.

We consider a fairly general class of potentials with self-similar elliptical equipotential and unit `core' radius of the form \cite{Bi,eva}
\be
\label{pota}
V (x,y;q,\alpha) =\cases { & $ {{1} \over {\alpha}} \left(1 + x^2 + \frac{y^2}{q^2}\right)^{\alpha/2}, \;\; \alpha \ne 0, $\cr
& $\frac12 \log\left(1 + x^2 + \frac{y^2}{q^2}\right), \;\;  \alpha = 0.$\cr}\ee
The ellipticity of the equipotentials is determined by the parameter $q$: we have an `oblate' figure when $q < 1$ and a `prolate' figure when $q > 1$. The range of $q$ is constrained by the condition that the density contours of the mass distribution generating the potential resembles real galaxies. If the slope parameter is restricted to the range
\be\label{alfa}
-1 < \alpha \le 2, \ee
Evans \cite{eva} shows that the allowed range for $q$ is an interval around 1 whose extent depends on $\alpha$ and shrinks to zero if $\alpha$ tends to $-1$. For example, in the logarithmic case $(\alpha=0)$ the range to get bulges or ellipticals is
\be\label{qrangez}
0.7 < q < 1.1\ee 
whereas, for $\alpha$ tending to $+1$ the range tends to the interval
\be\label{qrange1}
0 < q < 1.25.\ee

The family of potentials  (\ref{pota}) can be expanded in a series of the form (\ref{potes}). In view of the reflection symmetry with respect to both axes, the non-vanishing coefficients of the expansion up to order $N=4$ are the following:
\ba
C_{(2,0)} = & \frac12 \omega_1^2 = \frac12, &\label{q20} \\
C_{(0,2)} = & \frac12 \omega_2^2 = \frac1{2 q^{2}}, &\label{q02} \\ 
C_{(4,0)} = & \frac{\alpha-2}{8}, &\label{q40} \\
C_{(2,2)} = & \frac{\alpha-2}{4 q^{2}}, &\label{q22} \\
C_{(0,4)} = & \frac{\alpha-2}{8 q^{4}}. &\label{q04} \ea
The normal form truncated to the first non zero resonant term is given by expression (\ref{Kq11}) where, as can be verified by using the procedure above, $\delta = q - 1$ and the coefficients are given by
\ba
a & = & \frac34 \frac{C_{(4,0)}}{C_{(2,0)} \sqrt{2 C_{(0,2)}}} = \frac3{16} q (\alpha-2),\\
b & = & \frac34 \frac{C_{(0,4)}}{C_{(0,2)} \sqrt{2 C_{(0,2)}}} = \frac3{16 q} (\alpha-2),\\ 
c & = & d = \frac14 \frac{C_{(2,2)}}{C_{(0,2)} \sqrt{2 C_{(2,0)}  }} = \frac18 (\alpha-2).\label{11dsc}\ea

The model system given by (\ref{pota}) represents motion in the symmetry planes of a triaxial galaxy. In each of those planes, the symmetry axes directly give periodic orbits: at low energy, since the dynamics are slightly different from those of a harmonic oscillator ($\alpha = 2$), we may expect them to be stable oscillations. What happens when energy increases? Nonlinear dynamics give asynchronous motions, with frequencies depending on amplitudes. Instability can be triggered by low-order resonance and we can see how the normal form (\ref{Kq11}) is ideally suited to investigate this phenomenon.

By introducing the canonical variables adapted to the resonance by the resonant combinations
\be\label{calr11}
{\cal R}=J_1-J_2, \quad \psi=2 (\theta_{1}- \theta_{2}),\ee
the new Hamiltonian can be expressed in the {\it reduced} form 
\be\label{KER11}
K =  {\cal E} + 
\frac12 \delta ({\cal E} + {\cal R}) + A({\cal E}^{2} + {\cal R}^{2}) + B{\cal E}{\cal R} + C ({\cal E}^{2} - {\cal R}^{2})  (2 + \cos \psi), \ee
with 
\be
A = \frac14 (a+b)  , \quad
B = \frac12 (a-b) , \quad C = \frac{c}4 \label{ABC}\ee
and 
\be\label{cale11}
{\cal E}=J_1+J_2\ee
is the second integral as in (\ref{cale}). Considering the dynamics at a fixed value of ${\cal E}$, we have that $K$ defines a one--degree of freedom $(\psi,{\cal R})$ system with the following equations of motion
\ba
{\dot \psi} &=& 
\frac12 \delta + B{\cal E} + 2 \left(A - C (2 + \cos \psi)\right){\cal R},\label{dpsi1}\\
{\dot {\cal R}} &=& 
C \left({\cal E}^{2}  - {\cal R}^{2} \right) \sin \psi.\label{dr1}
\ea
The fixed points of this system give the periodic orbits of the original system. The pair of fixed points with ${\cal R}=\pm {\cal E}$ correspond to the normal modes: the orbit ${\cal R}={\cal E}$ (namely $J_{2}=0)$ is the periodic orbit along the $x$-axis (long axial orbit if $q<1$) whereas the orbit ${\cal R}=-{\cal E}$ $(J_{1}=0)$, is the periodic orbit along the $y$-axis (short axial orbit if $q<1$). 

However, we can see that additional periodic orbits may appear. These periodic orbits `in general position' exist only above a given threshold when the axial orbits change their stability. This phenomenon can be seen as a {\it bifurcation} of the new family from the normal mode when it enters in 1:1 resonance with a normal perturbation. The phase between the two oscillations also plays a role: these additional periodic orbits are respectively given by the conditions $\psi = 0$ ({\it inclined} orbits) and $\psi = \pm \pi$ ({\it loop} orbits). These conditions are solutions of ${\dot {\cal R}}=0$ (when ${\cal R}\ne\pm{\cal E}$) and determine the corresponding solutions of ${\dot \psi} = 0$:
\be\label{PO11A}
\psi = 0: \quad {\cal R}_{i}=\frac{\delta + 2B{\cal E}}{4(3C-A)}=\frac{\delta + (a-b){\cal E}}{3c-a-b}\ee
and
\be\label{PO11L}
\psi = \pm \pi: \quad {\cal R}_{\ell}=\frac{\delta + 2B{\cal E}}{4(C-A)}=\frac{\delta + (a-b){\cal E}}{c-a-b}.\ee
In view of (\ref{calr11}) and  (\ref{cale11}), the constraints $ 0 \le J_1,J_2\le{\cal E}$ 
applied to these solutions translate into the conditions of existence
\be\label{EPOA}
{\cal E} \ge {\cal E}_{i 1} = \frac{\delta}{3c-2a} \quad {\rm or} \quad  
{\cal E} \ge {\cal E}_{i 2} = \frac{\delta}{2b-3c} \ee
and 
\be\label{EPOL}
{\cal E} \ge {\cal E}_{\ell 1} = \frac{\delta}{c-2a} \quad {\rm or} \quad  
{\cal E} \ge {\cal E}_{\ell 2} = \frac{\delta}{2b-c}. \ee
Index $\ell$ denotes the critical values at which the loop orbits bifurcate (there are {\it two} of them, one clockwise, the other counter-clockwise) respectively from the $x$-axial normal mode if the first of (\ref{EPOL}) is satisfied and from the $y$-axial normal mode if the second one is satisfied. Index $i$ denotes the critical values at which the inclined orbits bifurcate, respectively from the $x$-axial normal mode if the first of (\ref{EPOA}) is satisfied and from the $y$-axial normal mode if the second one is satisfied. At the same bifurcation values, the normal mode suffers a change of stability, passing from stable to unstable when the new orbit is born. Also a second transition is possible (unstable to stable) if a second bifurcation ensues. 

\begin{figure}[h!]
\centering
\includegraphics[width=0.49\columnwidth]{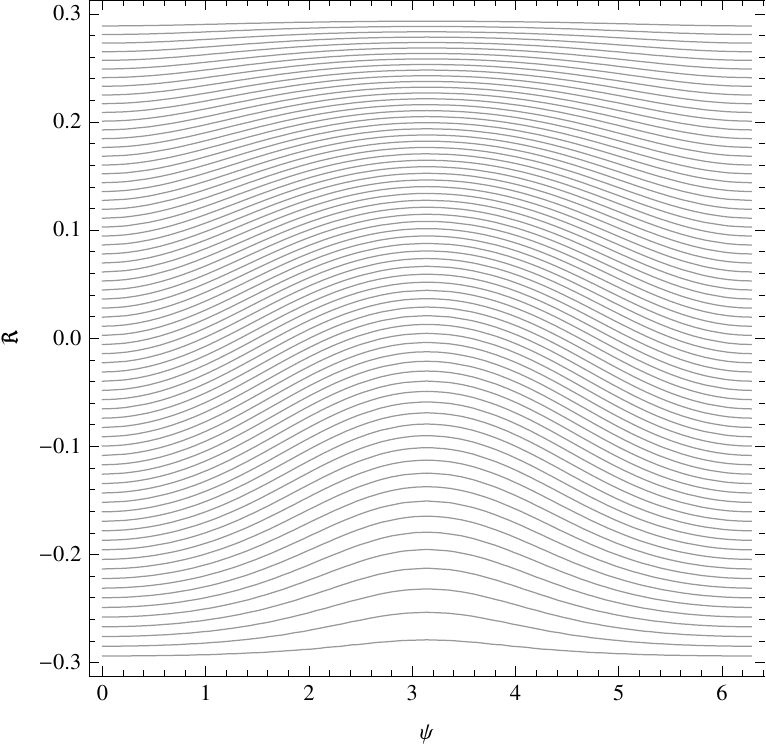}
\includegraphics[width=0.49\columnwidth]{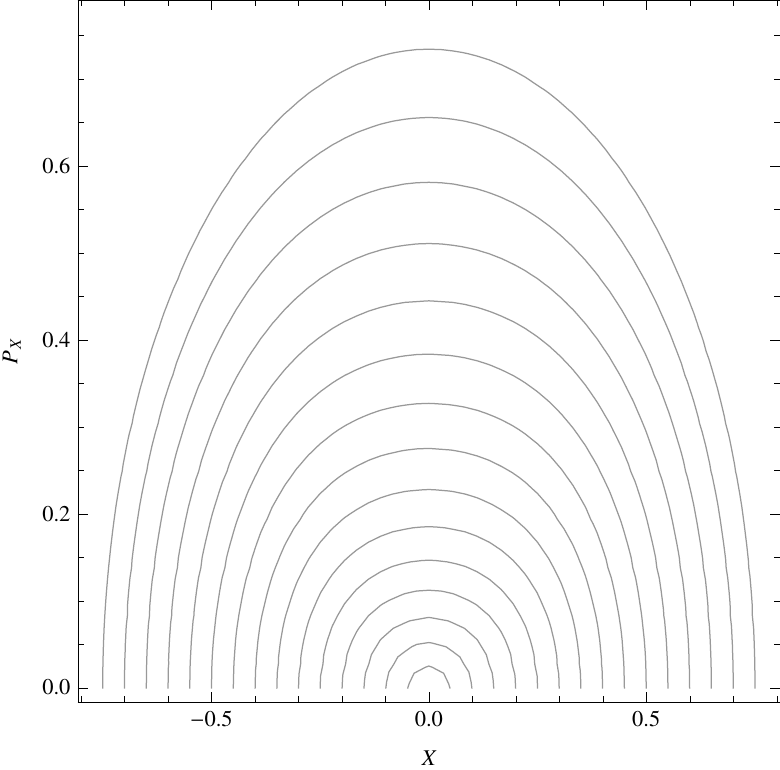}
\includegraphics[width=0.49\columnwidth]{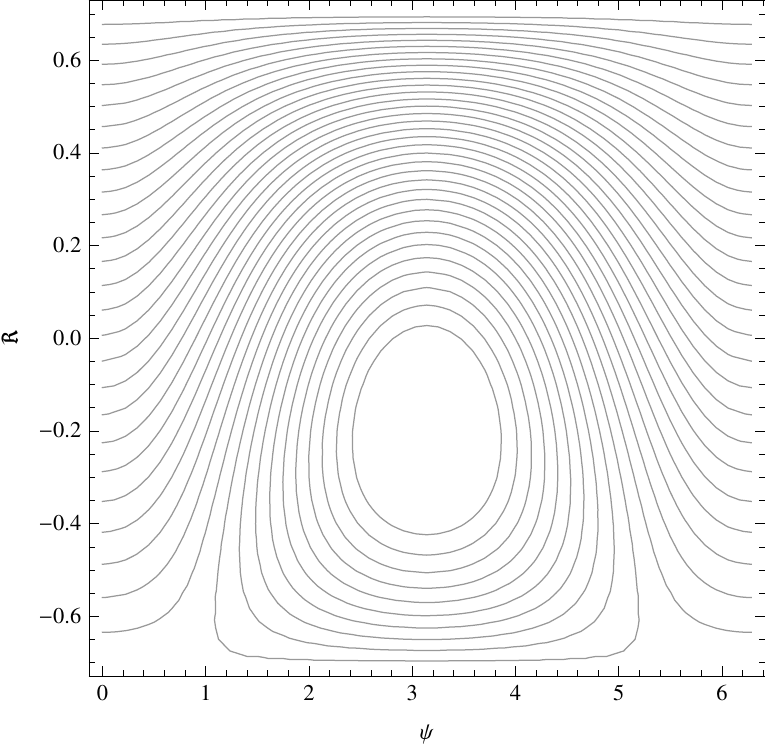}
\includegraphics[width=0.49\columnwidth]{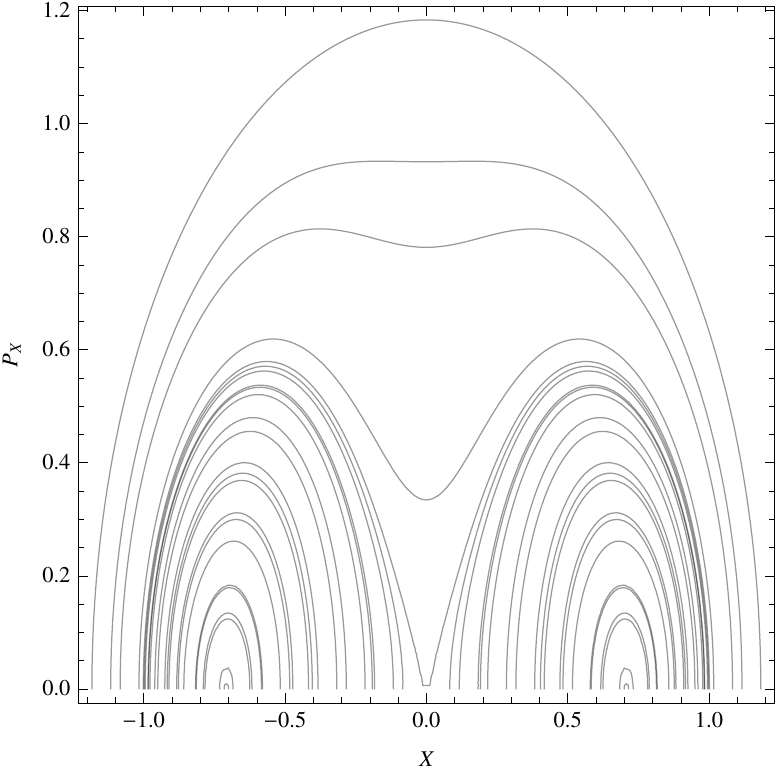}

\caption{Dynamics of the 1:1 doubly-symmetric resonant Hamiltonian, see text for explanation.}
\label{11q}
\end{figure}

In fig.\ref{11q} we can see the dynamics of the 1:1 doubly-symmetric resonant Hamiltonian displayed with the two coordinate sets exploited above: the left panels contain the level lines of the reduced Hamiltonian in the $(\psi,{\cal R})$ cylinder with the lines $\psi = 0$ and $\psi = 2\pi$ identified; the right panels are the $(X,P_X)$ `surfaces of section' computed with the condition $Y=0, P_Y > 0$. We have chosen a model with an oblate ($q = 0.7$) `logarithmic' potential ($\alpha = 0$).
The upper panels are computed for ${\cal E} = 0.3$, that is below ${\cal E}_{\ell 2} (=0.365)$: on the left we see the continuous curves associated to the `rotational' invariant tori around the two stable normal modes; on the right we see a section of these tori corresponding to `box' orbits around the minor-axis. The lower panels are computed for ${\cal E} = 0.7$, that is above ${\cal E}_{\ell 2}$: on the left we see that at $\psi = \pi$ a stable fixed point has appeared surrounded by `librational' invariant tori; on the right we see that the minor-axis is now unstable and there are {\it two} stable fixed points corresponding to the two periodic loops orbiting in the two senses of rotation. Each of them is surrounded by non-periodic loops. There is also a {\it separatrix} beyond which there are the remaining boxes parented by the mayor-axis which is still stable.

Fig.\ref{11q} nicely provides qualitative informations. We stress that these results are also {\it quantitatively} useful. In view of the rescaling and of the expansion of the energy as a truncated series in the parameter ${\cal E}$, we have that $E = {\omega_2} {\cal E} = {\cal E} / q$ is a first order estimate of the `true' energy of the orbital motion. We can use the above critical values to establish the instability threshold for the model problem given by potentials (\ref{pota}). For the parameter ranges used in galactic dynamics, it happens that the typical bifurcation is that of the loop orbits at the critical energy
\be\label{DS11L}
E_{\ell} = \frac{4|q-1|}{2-\alpha}.\ee
In the range
\be
0.7 < q < 1.3,\ee
which can be considered as `realistic' for elliptical galaxies, the thresholds (\ref{DS11L}) give estimates correct within a 10\% if compared to numerical computations \cite{bbp,pbb}.

\section{Resonances and bifurcations in axisymmetric potentials}

Here we discuss how a slight generalization of the normal form studied above is able to describe some very important features of the dynamics of {\it axisymmetric} galaxies. A spheroidal bulge or a disk/halo system can be modeled with a potential with an exact axial symmetry and the conservation of the angular momentum $L$ with respect to the symmetry axis still allows us to stay with only two degrees of freedom. The dynamics is then given by the family of potentials
\be\label{pote}
V_{\rm eff} (R,z; L,q, \alpha) = \frac{L^{2}}{2{R}^{2}} + V (R,z;q,\alpha),\ee
where potentials of the same form as in (\ref{pota}) can be used by exploiting cylindrical coordinates $R,\phi,z$.
These {\it effective} potentials have a unique absolute minimum in 
\be\label{amin}
R=R_{\rm c}(\alpha)=L^{\frac2{2+\alpha}},\;\;z=0.\ee
This is a stable equilibrium on any {\it meridional} plane $\phi = {\rm const}$ corresponding to a circular orbit of radius $R_{\rm c}(\alpha)$ of the full three-dimensional problem. Now we have four parameters, $L,q, \alpha$ and the energy to take into account. We can get rid off of the energy by putting to zero the core radius so that the dynamics are scale-free and we may fix energy at some convenient value:
\be\label{ena}
E \doteq E_{\alpha} = \cases { &$ \left(\frac12 + \frac{1}{\alpha}\right){\rm e}^{-\frac{\alpha}{2+\alpha}}, \;\; \alpha \ne 0, $\cr
&$ 0, \;\;  \alpha = 0.$\cr}\ee
This implies that the radius of the circular orbit at this energy is
\be\label{circ}
R_{\rm c}(\alpha)={\rm e}^{-\frac{1}{2+\alpha}}, \;\; -1 < \alpha \le 2\ee
and we can investigate the dynamics at 
$E=E_{\alpha} $ 
by varying $L$ in the range
\be\label{amr}
0 < L \le L_{\rm max} \equiv R_{\rm c}^{\frac{2+\alpha}2} = \frac1{\sqrt{\rm e}}\ee
without any loss of generality \cite{hu,p09}.

In order to implement the perturbation method we expand the effective potential around the minimum (\ref{amin}). We introduce rescaled coordinates according to
\be\label{xy}
x \doteq \frac{R-R_{\rm c}}{R_{\rm c}}, \quad y \doteq \frac{z}{R_{\rm c}}\ee
with origin in the equilibrium point (\ref{amin}). The potential  (\ref{pote}) is expanded as a truncated series (in the coordinates $x,y$) of the form (\ref{potes}),
where the truncation order $N$ is determined by the resonance under study: here we limit ourselves to the 1:1 case. From (\ref{amin}) and the rescaling (\ref{xy}), the constant term of the expansion is
\be\label{CZZ}
C_{(0,0)} (L,\alpha) =\cases { & $ \left(\frac12 + \frac{1}{\alpha}\right)L^{\frac{2\alpha}{2+\alpha}}, \;\; \alpha \ne 0, $\cr
& $\frac12 + L, \;\;  \alpha = 0$\cr}\ee
and the other coefficients have the form
\be\label{CJK}
C_{(j,k-j)} (L,\alpha,q) = L^{\frac{2\alpha}{2+\alpha}} c_{(j,k-j)} (\alpha,q).
\ee
In order to simplify formulas, we introduce the new parameter
\be\label{beta}
\beta = -\frac{2\alpha}{2+\alpha}, \;\; -1 < \beta \le 2, \ee
with the same range of $\alpha$ in view of (\ref{alfa}).

The orbit structure of the original family of potentials (\ref{pote}) at the energy level fixed by (\ref{ena}) will be approximated by the orbit structure of the rescaled Hamiltonian
\be\label{HaSa}
\widetilde H = \frac12 \big(p_{x}^{2} + p_{y}^{2} \big) + \widetilde V_{\rm eff} (x,y;q,\alpha),\ee
where
\be\label{pots}
\widetilde V_{\rm eff} (x,y;q,\alpha) = L^{\beta} V_{\rm eff} (x,y;L,q,\alpha).\ee
The dynamics given by Hamiltonian (\ref{HaSa}) take place in the rescaled time
\be\label{TS}
\tau = t/L^{\beta+1}\ee
at the new `energy'
\be\label{EL}
E_F = L^{\beta} \left(E_{\alpha} - C_{(0,0)} (L,\alpha) \right) = 
  \frac1{\beta} \left(1-(L/L_{\rm max})^{\beta} \right).\ee
According to (\ref{amr}) the singular value $L=0$ is excluded from the analysis implying that the fictitious energy (\ref{EL}) is always finite and that the expansion around the equilibrium point (\ref{amin}) make sense. The equilibrium value $E_F=0$ corresponds to the circular orbit with $L=L_{\rm max}$ and lower values of $L$ give increasing values of $E_F$.

The non-vanishing coefficients of the expansion of $\widetilde V_{\rm eff}$ up to order $N=4$ are the following:
\ba
c_{(2,0)} = & \frac12 \omega_1^2 = \frac{2+\alpha}2, &\label{c20} \\
c_{(0,2)} = & \frac12 \omega_2^2 =\frac1{2 q^{2}}, &\label{c02} \\ 
c_{(3,0)} = & -\frac{10+3 \alpha-\alpha^{2}}6, &\label{c30} \\
c_{(1,2)} = & -\frac{2-\alpha}{2 q^{2}}, &\label{c12} \\
c_{(4,0)} = & -\frac{54+11 \alpha-6\alpha^{2}+\alpha^{3}}{24}, &\label{c40} \\
c_{(2,2)} = & -\frac{6-5 \alpha-\alpha^{2}}{4 q^{2}}, &\label{c22} \\
c_{(0,4)} = & -\frac{2-\alpha}{8 q^{4}}. &\label{c04} \ea
The first two of them provide the frequencies of the {\it epicyclic motions}. Recalling (\ref{uf}) and the time rescaling in (\ref{TS}), the radial and vertical harmonic frequencies are respectively
\be\label{rf}
\kappa = \frac{\omega_1}{ L^{\beta+1}} = \frac{\sqrt{2+\alpha} }{ L^{\beta+1}} \ee
and
\be\label{vf}
\nu = \frac{\omega_2}{ L^{\beta+1}} = \frac{1}{q L^{\beta+1}}. \ee
Therefore, to describe the bifurcations ensuing from the 1:1 resonances, the normal form is computed by using the detuning parameter defined through 
\be\label{DET11}
\delta = \frac{\kappa}{\nu} - 1 = q \sqrt{2+\alpha} - 1. \ee

The presence of coefficients of cubic terms in the expansion accounts for the breaking of the reflection symmetry with respect to the $y$ axis: we still have the reflection symmetry with respect to the equatorial plane corresponding to the $x$-axis in the coordinates (\ref{xy}). The $X$-axis ($J_1$) normal mode corresponds, in the full 3-dimensional problem, to the equatorial non-periodic disk orbits. Breaking the symmetry with respect to the `vertical axis' means that the $y$-normal mode corresponds to another orbit that in the 3-dimensional space is confined to a folded surface: the so called {\it thin tube}. This is an example in which we see that normalizing transformation are, in a sense, `rectifying' transformations. On this respect, it is also useful to recall something about the taxonomy of periodic orbits. Again we denote the bifurcating family as the {\it inclined} orbit in view of its natural interpretation as the {\it in phase} ($\psi=0$) 1:1 resonance of the two oscillations \cite{bbp,zm}. The anti-phase resonant loops never appear in these systems, at least as a stable family (see below). The inclined periodic orbits parent two families of inclined boxes that may arrive quite far from the equatorial plane both above and below the disk: this phenomenon is called {\it levitation} \cite{ST1}. We recall that our inclined orbits have also been referred to as {\it reflected banana} by \cite{LS} and \cite{eva} and simply as {\it banana} by \cite{hu}: we prefer to leave this term as the standard denomination \cite{mes} for the 2:1 resonance. They exist also nicknames for high-resonant periodic orbits ({\it fish, pretzel, etc.} \cite{mes,hu}).

The normal form truncated to the first non zero resonant term is given by the general expression (\ref{Kq11}) where the coefficients are
\ba
a & = & \frac32 \frac{c_{(4,0)}}{\omega_1^2 \omega_2} - \frac{15}4 \frac{c_{(3,0)}^2}{\omega_1^3 \omega_2^2}
      = \frac{q (-5 q (5 - \alpha)^2 (2 + \alpha) + 3 \sqrt{2 +\alpha} (27 - 8 \alpha + \alpha^2))}{48 \sqrt{2 + \alpha}}, \label{11acq}\\
b & = & \frac32 \frac{c_{(0,4)}}{\omega_2^3} - \frac5{12} \frac{c_{(1,2)}^2}{\omega_1 \omega_2^4} 
      = \frac1{48} (\alpha-2) \left(\frac9{q} + \frac{5 (2 - \alpha)}{\sqrt{2 + \alpha}} \right),\\ 
c & = & \frac{c_{(2,2)}}{2 \omega_1 \omega_2^2} - \frac{3 c_{(1,2)} c_{(3,0)}}{2 \omega_1^2 \omega_2^3} - 
             \frac{c_{(1,2)}^2}{3 \omega_1 \omega_2^4}
      = \frac{(10 - 7 \alpha + \alpha^2) (1 - 3 q \sqrt{2 + \alpha})}{24 \sqrt{2 + \alpha}},\\
d & = & \frac{c_{(2,2)}}{2 \omega_1 \omega_2^2} + \frac{c_{(1,2)} c_{(3,0)}}{2 \omega_1^2 \omega_2^3} - 
             \frac{c_{(1,2)}^2}{\omega_1 \omega_2^4}
      = \frac{(\alpha-2) (3 - 5 q \sqrt{2 + \alpha} - \alpha (3 - q \sqrt{2 + \alpha})}{24 \sqrt{2 + \alpha}}. \label{11dcq}\ea
Comparing with eq.(\ref{11dsc}) we see that the effect of the symmetry breaking is that now $d \ne c$ with more complex bifurcation conditions. 

By lowering the angular momentum (namely by increasing the fictitious energy $E_F$) both normal modes may lose their stability through a 1:1 resonance. The equations of motion for the variables adapted to the resonance now are
\ba
{\dot \psi} &=& {\widetilde K}_{\cal R} = 
\frac12 \delta + B{\cal E} + \frac12 \left(4 A - (2c + d  \cos \psi)\right){\cal R},\label{dpsi2}\\
{\dot {\cal R}} &=& - {\widetilde K}_{\psi} = 
\frac{d}4 \left({\cal E}^{2}  - {\cal R}^{2} \right) \sin \psi.\label{dr2}
\ea
The analysis of the fixed points of this system proceed as in the previous section. The pair of fixed points with ${\cal R}=\pm {\cal E}$ correspond to the normal modes. The periodic orbits in `general position' (inclined and loops) are respectively given by the solutions:
\be\label{POcqA}
\psi = 0: \quad {\cal R}_{i}=\frac{\delta + 2B{\cal E}}{2c + d - 4A}=\frac{\delta + (a-b){\cal E}}{2c - a - b + d}
\ee
and
\be\label{POcqL}
\psi = \pm \pi: \quad {\cal R}_{\ell}=\frac{\delta + 2B{\cal E}}{2c - d - 4A}=\frac{\delta + (a-b){\cal E}}{2c - a - b - d}.\ee
The constraints $ 0 \le J_1,J_2\le{\cal E} $ 
applied to these solutions translate into the conditions of existence
\be\label{EPOcqA}
{\cal E} \ge {\cal E}_{i 1} = \frac{\delta}{2(c-a)+d} \quad {\rm or} \quad  
{\cal E} \ge {\cal E}_{i 2} = \frac{\delta}{2(b-c)-d} \ee
and  
\be\label{EPOcqL}
{\cal E} \ge {\cal E}_{\ell 1} = \frac{\delta}{2 (c - a) - d} \quad {\rm or} \quad  
{\cal E} \ge {\cal E}_{\ell 2} = \frac{\delta}{2 (b - c) + d}.\ee
By using (\ref{11acq}--\ref{11dcq}) the bifurcation equations (\ref{EPOcqA},\ref{EPOcqL}) determine the {\it critical} values of ${\cal E}$ in terms of the parameters $q,\alpha$. A common situation, for models ranging from sensibly oblate to prolate and with $\alpha$ in the range (\ref{alfa}), is that in which the $x$-axis becomes unstable and the inclined appears as a bifurcation from the disk orbit. The passage to instability of the $y$-axis is possible for oblate models such that 
\be\label{qcr}
q < \frac1{\sqrt{2 +\alpha}}\ee 
and gives rise to a bifurcation from the thin tube. In this case, also the second bifurcation is usually possible: the disk orbit returns stable and the new family (loops) is unstable. In fig.\ref{11cq} we can see the dynamics of the 1:1 resonant Hamiltonian with one of the symmetries broken, displayed with the same settings as in fig.\ref{11q}. We have chosen a model with an oblate `logarithmic' potential ($\alpha = 0$) with a value of $q=0.65$ in order to satisfy inequality (\ref{qcr}) to show the occurrence of both bifurcations. 
The upper panels are computed for ${\cal E} = 0.13$, that is above ${\cal E}_{i 2} (=0.10)$: on the left we see the island at  $\psi = 0, 2\pi$ around the stable inclined orbits bifurcated from the thin tube; on the right we see a section of the tori around the stable inclined orbit, a separatrix departing from the thin tube (the unstable normal mode) and the tori beyond it associated to the box orbits parented by the stable disk. The lower panels are computed at ${\cal E} = 0.15$, that is above ${\cal E}_{\ell 2} (=0.135)$, the threshold of the second bifurcation: on the left we see that at $\psi = \pi$ an unstable fixed point has appeared, testifying the instability of the loops ensuing from this second bifurcation; on the right we see that the the thin tube is again stable and there are {\it two} unstable fixed points corresponding to the two loops. The thin tube is now surrounded by tori associated to the {\it thick} tubes and there the other two families of tori around the inclined and around the disk.

\begin{figure}[h!]
\centering
\includegraphics[width=0.49\columnwidth]{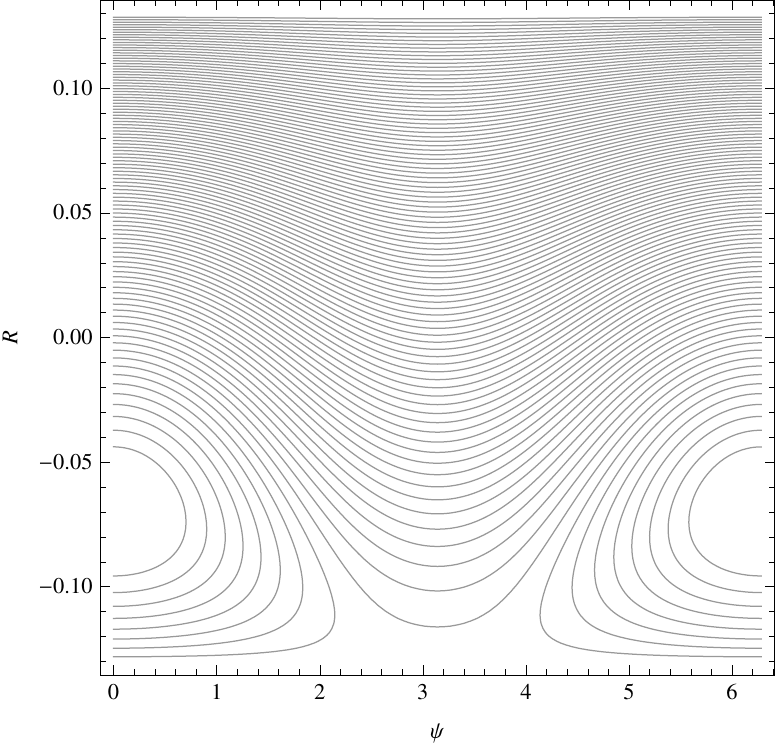}
\includegraphics[width=0.48\columnwidth]{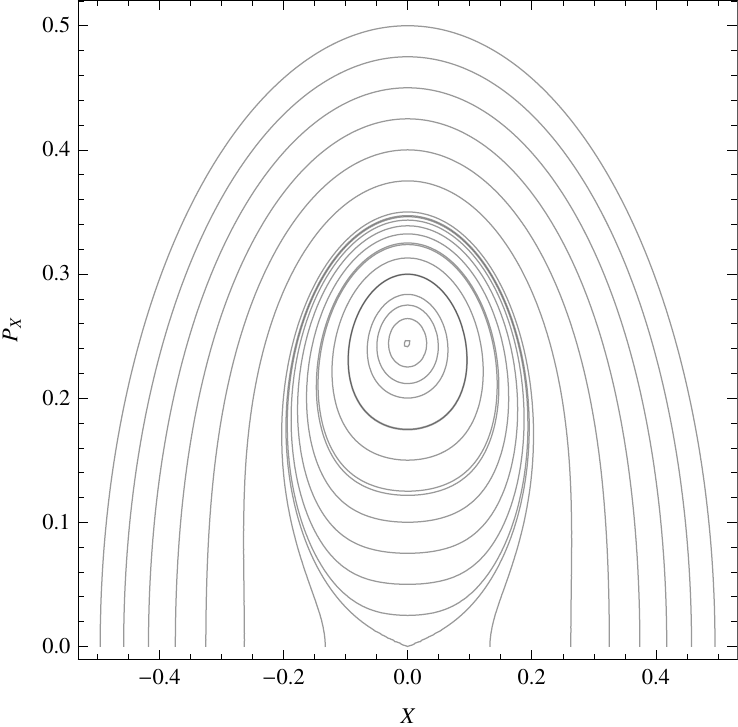}
\includegraphics[width=0.49\columnwidth]{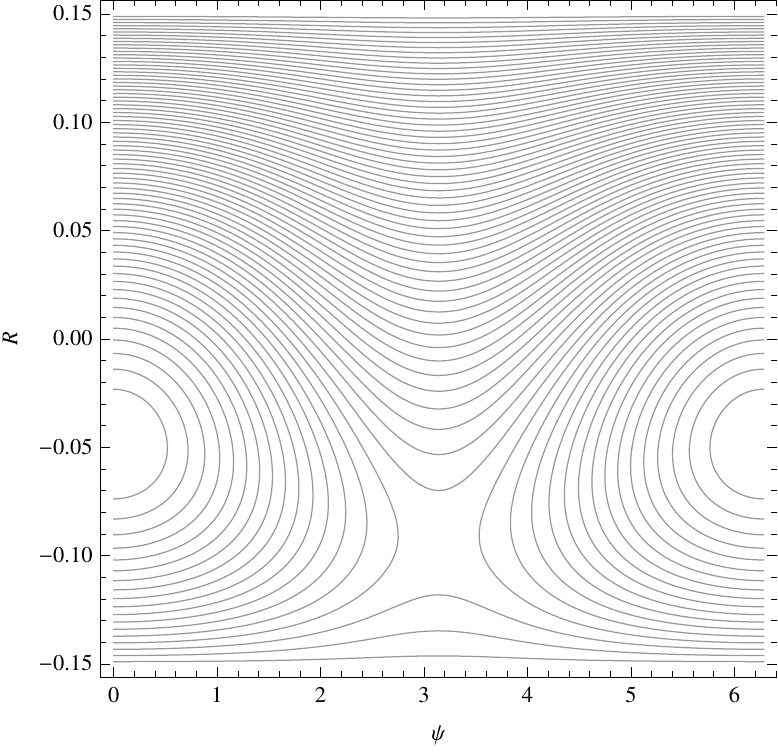}
\includegraphics[width=0.48\columnwidth]{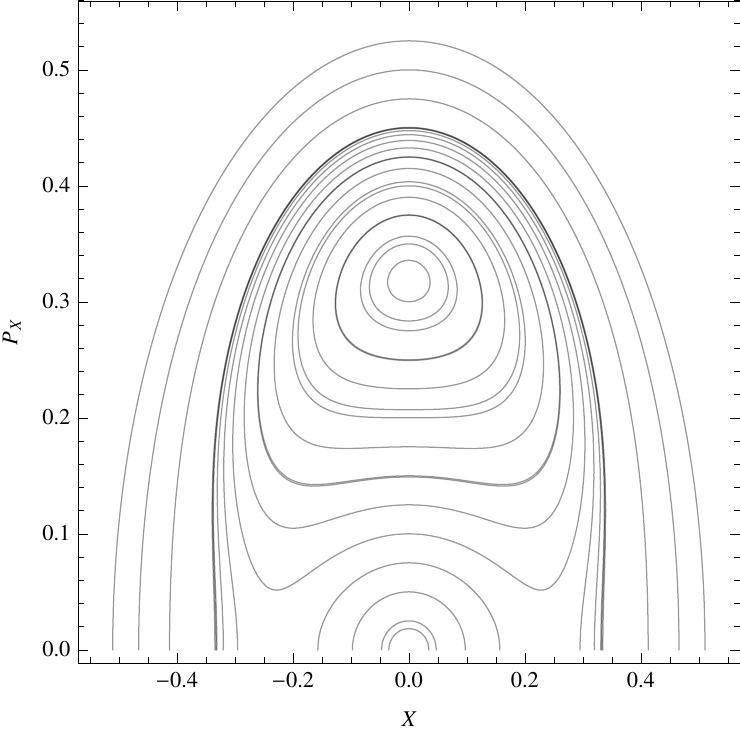}

\caption{Dynamics of the 1:1 simply-symmetric resonant Hamiltonian, see text for explanation.}
\label{11cq}
\end{figure}

To make quantitative predictions, we want expressions for the corresponding {\it critical angular momentum}. The approach we have followed so far is altogether a perturbation approach truncated to the first non-trivial order: therefore it is natural to look also in this case for expansions truncated to the first order in the detuning parameter. Taking into account the rescaling in $\widetilde K $, conditions (\ref{EPOcqA}) and the explicit expressions (\ref{11acq}--\ref{11dcq}), the first order expansions of the critical values of the fictitious energy (\ref{EL}) are:
\be\label{c11}
E_F = \cases { & $ \frac{12 (2+\alpha)}{5(-2-\alpha+\alpha^{2})}  \delta, \quad  \delta < 0,$\cr
                                    & $ \frac{6 (2+\alpha)}{2+\alpha-\alpha^{2}}  \delta, \quad  \delta > 0.$\cr}\ee
                                    The first solution corresponds to the bifurcation from the thin tube, the second one corresponds to the bifurcation from the disk. These are again examples of series of the form (\ref{detexp}) truncated to the first order.
                                   Afterwards, using the relation between $E_F$ and $L$ established by (\ref{EL}), we get the following expressions for the critical values of the angular momentum below which inclined orbits exist:
\be\label{Lc11}
L_{\rm crit} = \frac1{\sqrt{\rm e}} \times \cases { 
& $  \left(1- \frac{24 \alpha  (q \sqrt{2+\alpha} - 1)}{5(2+\alpha-\alpha^{2})} \right)^{-\frac{2+\alpha}{2\alpha}}, 
\quad  q < \frac1{\sqrt{2+\alpha}},$\cr
& $  \left(1+ \frac{12 \alpha (q \sqrt{2+\alpha} - 1)}{2+\alpha-\alpha^{2}} \right)^{-\frac{2+\alpha}{2\alpha}}, 
\quad  q > \frac1{\sqrt{2+\alpha}}.$\cr}\ee
It is also useful to write the limiting case of the logarithmic potential ($\alpha=0$):
\be\label{LcL11}
L_{\rm crit} = \cases { & $  {\rm e}^{-\frac{29}{10} + \frac{12}{5} \sqrt{2} q}, \quad  q < \frac1{\sqrt{2}},$\cr
                                   & $ {\rm e}^{\frac{11}2 - 6 \sqrt{2} q}, \quad  q > \frac1{\sqrt{2}}.$\cr}\ee 
A comparison with the outcome of numerical determinations of the bifurcation threshold allows us to evaluate the accuracy of these analytical predictions. In \cite{p09} the critical value of the angular momentum for the bifurcation of the inclined orbits, computed with Eq.(\ref{Lc11}) for general $\alpha$ and with eq.(\ref{LcL11}) for $\alpha=0$, are compared with numerical data obtained either from published works \cite{LS,eva,hu} or by numerical computations made specifically for that paper. In this case, the bifurcation has been detected tracing the instability threshold of the normal mode by means of the Floquet method. The accuracy is particularly good when the model is close to the exact resonance. Overall, the discrepancy linearly grows with detuning, as can be expected in this first order approach. In the case of two strongly oblate models, the thin tube becomes unstable: in the rather extreme case with $q=0.4,\alpha=0.5$, that is just at the margin of the physical range established in \cite{eva}, the detuning is $\delta=-0.37$ and the relative error in the prediction is 18\%. In all other cases, the disk becomes unstable and the quality of the prediction can be represented by the cases with $q=0.8,\alpha=0.1$ studied by Hunter and collaborators \cite{hu}, where the detuning is $\delta=0.16$ and the relative error in the prediction is 8\% and $q=0.85,\alpha=-0.18$ which was investigated by Evans \cite{eva}, where the detuning is $\delta=0.14$ and the relative error in the prediction is 7\%. 

As said above, a second stability change may occur when the second resonant family appears \cite{mes}. In this setting, this possibility occurs if the loops appear. By using the bifurcation equations (\ref{EPOcqL}) and the explicit expressions (\ref{11acq}--\ref{11dcq}), the inequality can be satisfied only with values of the parameters corresponding to rather extreme oblate models. This implies a negative value of the detuning and a critical fictitious energy
\be\label{cl11}
E_F = -\frac{4 (2+\alpha)}{2+\alpha-\alpha^{2}}  \delta.\ee
We get the following expression for the critical value of the angular momentum below which loops bifurcate from the thin tube:
\be\label{Lcl11}
L_{\rm crit} = \frac1{\sqrt{\rm e}} \left(1- \frac{8 \alpha  (q \sqrt{2+\alpha} - 1)}{2+\alpha-\alpha^{2}} \right)^{-\frac{2+\alpha}{2\alpha}}. \ee                     
   This is again a `pitchfork' bifurcation so that the thin tube regains its stability. The loops are unstable and, lowering the angular momentum below the critical value, remain unstable for every reasonable combinations of the parameters. At the same time, the inclined tend to occupy an even larger fraction of phase space. We are now very far from the harmonic behaviour, therefore it is reasonable to expect a certain discrepancy between the prediction and the actual value. Referring again to the models in \cite{p09}, in the case with $q=0.4,\alpha=0.5$ the `true' critical value for the return to stability is $0.32$ \cite{hu} whereas (\ref{Lcl11}) predicts $0.18$. In the concluding remarks we give some hints on how to improve the quality of the predictions. 

\section{Discussion}

Our aim has been to show how to make useful predictions on the dynamics with the `tool-box' offered by the normal-form theory. The reader can also simply choose one of these tools and use it in some specific problem (if it can be fitted to the classes above) even without a deep involvement into the theory. However, we have also tried to lay out a sketch of how to proceed from the oustart. We remark that this approach is not the only possible one and that many others have been devised: after all, many of the phenomena described above are a form of nonlinear resonance in mechanics, a subject to which many efforts have been devoted \cite{landau} in the long history of this field. 
However, to testify the relevance of the normal-form method, it has also been applied as a tool to investigate parametric resonance in its more general mathematical formulation \cite{bs}.

In order to improve the predictive power of the approach, the standard way to proceed is that of going to higher orders in the expansions. Following the procedure illustrated in section 2, an algorithm can be devised to construct a normal form up to arbitrary order. With this, one can compute the bifurcation thresholds as series in the detuning truncated at an order to be determined by the required precision. Very good predictions can be performed with series truncated already at degree 6--8 \cite{bbp,p08} in the phase-space variables for several kinds of resonance. Clearly, an immediate warning to do is that the procedure may result in cumbersome formulas, therefore some educated exploitations of automatic computations is mandatory. But there is also a more subtle issue with high-order expansions: the series we are working with are in general not converging. However, they are nonetheless asymptotic series. This means that there exist an `optimal' order at which to stop anyway the expansions because beyond it nothing is gained in precision (for a discussion see ref.\cite{pbb}). There is in general no way to know the optimal order in advance, however experience with known systems and comparison with numerical analysis suggests that, when/if the system enters a chaotic regime (e.g. when energy overcomes a certain stochasticity threshold \cite{HH}) the optimal order apt to describe the dynamics near to the chaotic transition is just the order at which resonant terms firstly appear in the normal form. 

As far as the applications in galactic dynamics, there are several instances in which some progresses can most probably be done with resonant normal forms. We mention the study of rotating systems in which not much more than the results obtained in the pioneering works of de Zeeuw and Merritt \cite{zm} is available from the analytical side and the investigation has recently be done only numerically. 
Another issue of great relevance is that of a central black hole (or some other kind of gravitational singularity like in a strong galactic cusp) in which the dynamics are that of a quasi-Keplerian perturbed field. Regularizing coordinate transformations allows us to turn the system into a perturbed oscillator and the normal form method can again be applied. Finally, it is clear that an important step forward is that of investigating fully tri-dimensional systems. In this case, in general, the normalizing transformation no longer provides an integrable normal form. The gain is usually limited to decrease by one the dimensionality of the problem. However, the investigation of existence and stability of periodic orbits proceeds in much the same way as above, even if at the price of a higher algebraic complexity. This is probably the most promising area of future investigations.

\section*{Acknowledgments}
This work is supported by INFN -- Sezione di Roma Tor Vergata and by the Scuola di Dottorato of the Dipartimento di Scienze di Base e Applicate per l'Ingegneria -- Universit\`a di Roma ``la Sapienza".

\end{document}